\documentclass{article} 
\usepackage{iclr2027_conference,times}

\usepackage{amsmath,amsfonts,bm}

\def\eqref#1{equation~\ref{#1}}

\def\1{\bm{1}}

\DeclareMathAlphabet{\mathsfit}{\encodingdefault}{\sfdefault}{m}{sl}
\SetMathAlphabet{\mathsfit}{bold}{\encodingdefault}{\sfdefault}{bx}{n}

\usepackage{tikz}
\usetikzlibrary{
    arrows.meta,
    positioning,
    fit,
    calc,
    backgrounds
}
\usepackage{booktabs}
\usepackage{tabularx}
\usepackage{graphicx}
\usepackage{hyperref}
\usepackage{url}

\title{Intrinsic Goals for Autonomous Agents: \\Hierarchical Exploration Captures \\Future Volitional Action in Drosophila}

\author{
Declan Kutscher$^{1,2,}$\thanks{Equal contribution.}\,\,\,,
Reece Keller$^{3,4,*}$,
Alyn Kirsch$^{4}$,
Felix Pei$^{3,2}$ \&
Aran Nayebi$^{2,3,4}$ \\
$^1$Department of Computer Science, University of Maryland, College Park \\
$^2$Machine Learning Department, Carnegie Mellon University \\
$^3$Neuroscience Institute, Carnegie Mellon University \\
$^4$Robotics Institute, Carnegie Mellon University \\
\texttt{declank@umd.edu, \{rdkeller, akirscht, fcpei, anayebi\}@cs.cmu.edu}
}

\usepackage{xcolor}

\newcommand{\beforeturn}{$-15$ to $-8\,\mathrm{s}$}
\newcommand{\duringturn}{$-0.4$ to $+0.4\,\mathrm{s}$}

\iclrfinalcopy
\begin{document}

\maketitle
\begingroup
\renewcommand{\thefootnote}{}
\footnotetext[0]{Project website: \url{https://neuroagents-lab.github.io/autonomous_drosophila/}}
\endgroup
\begin{abstract}
Animals have a remarkable ability to act autonomously---driven by intrinsic motivation---without external sensory cues or extrinsic objectives, yet the computational mechanisms behind this ability largely remain a mystery. To understand how biological autonomy may emerge from optimizing intrinsic goals, we use recent whole-brain recordings in the fruit fly (\textit{Drosophila melanogaster}) that identify neural representations of spontaneous exploratory decisions more than $10$ seconds in advance. Combined with a neuromechanical digital twin of \textit{Drosophila}, these recordings allow us to test whether intrinsically motivated artificial agents with realistic embodiments quantitatively match fly behavior and internal representations of future volitional action during exploration. We implement a hierarchical, intrinsically motivated agent whose policy operates over a discrete library of motor primitives---fixed low-level controllers that actuate joint dynamics---closely matching ethological theories of behavioral syllables. We train agents with no external reward using $20$ intrinsic motivation algorithms spanning curiosity, entropy, and competency methods. We evaluate frozen policies in a virtual tethered-ball assay modeled after the fruit-fly experiment, allowing us to compare spontaneous locomotor statistics and internal representations and decode future turn direction from both kinematics and neural activations on the same timescales observed in flies.
We find that DIAYN---a hierarchical competency algorithm that learns a discrete set of policies maximizing behavioral entropy and the separability of their induced sensory observations---achieves the strongest joint behavioral and neural prospective coding performance with the fruit fly. 
Surprisingly, DIAYN's learned policies closely capture individual variability in fly exploration strategies, suggesting that this behavioral diversity optimizes a concrete computational goal of maximizing state-space coverage at the \emph{population} level. 
Furthermore, DIAYN agents' latent activity closely captures the specific sum-difference rule in interhemispheric encoding of future kinematics in biological flies.
Together, these results establish hierarchical, intrinsically motivated agents as a powerful framework for reverse-engineering animal autonomy.
\end{abstract}

\section{Introduction}
\begin{figure}[t!]
  \centering
  \includegraphics[width=\textwidth, page=1]{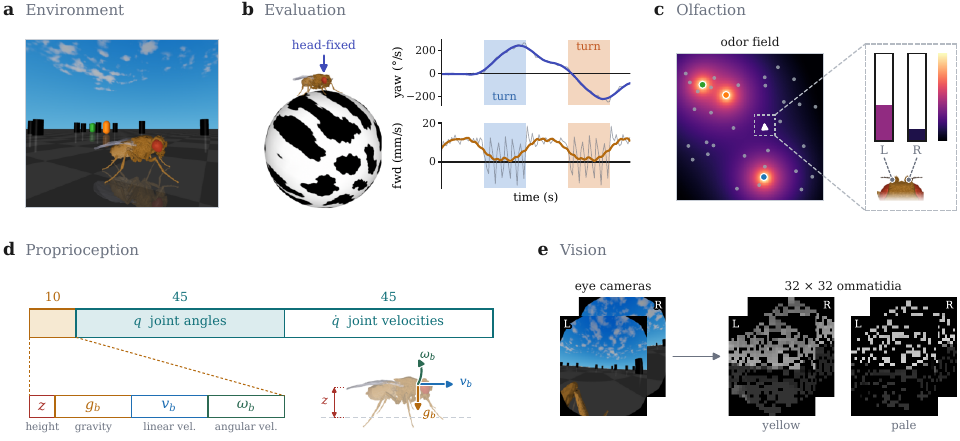}
  \caption{\textbf{Environment Overview.}
\textbf{(a)} The fly policy is trained in an environment with obstacles and multiple odor sources. 
\textbf{(b)} The policy is then evaluated in a tethered-ball environment where the fly is head-fixed. The angular and forward velocities are read from the ball, analogous to the biological experiment of \citet{brezovec2023volitional}.
\textbf{(c)} The policy receives olfaction as a weighted sum of intensities sampled at the antennae and maxillary palps. 
\textbf{(d)} The policy's proprioceptive input comes directly from NeuroMechFly and includes joint angles, joint velocities, and body-frame gravity, height, and linear and angular velocities.
\textbf{(e)} Vision is also provided to the policy by simulating the ommatidia mapping of pale and yellow type cells to pixels.
}
  \label{fig:fig_1_overview}
\end{figure}

Animals continuously structure experience even in the absence of an explicit task, external reward, or instructive sensory cue. They explore their environment and learn about how their own actions affect the world before those experiences are necessarily tied to a particular extrinsic goal. Such self-directed experience may contribute to the flexibility with which animals later learn new tasks, adapt to changing environments, and acquire new behaviors over their lifetime. In contrast, artificial agents are commonly optimized for extrinsic objectives and task-specific reward functions, and achieving similarly broad and persistent autonomy remains challenging \citep{hughes2024open}. Understanding the computational objectives that explain intrinsically-driven animal behavior may therefore provide principles for building artificial agents that autonomously learn useful representations and behaviors \emph{without} extrinsic goals \citep{keller2025intrinsic}. 

A wide range of intrinsic-motivation objectives have been proposed as computational mechanisms for generating autonomous behavior in the context of exploration and reinforcement learning. These include objectives such as novelty \citep{burda2019rnd}, prediction error \citep{pathak2017icm}, uncertainty \citep{pathak2019disag}, state-space coverage \citep{liu2021apt}, empowerment \citep{klyubin2008emppot}, and unsupervised skill discovery \citep{eysenbach2019diayn}. However, these methods are typically evaluated according to criteria designed for artificial agents, such as downstream task performance on highly specific environments. 

Such evaluations do not directly test for autonomy, nor do they establish whether the resulting behavior or internal representations resemble those of biological agents---systems that \emph{already} exhibit robust autonomous behavior. 
Biological systems therefore provide an additional and independently motivated constraint on the space of possible intrinsic objectives. 
Hence, a candidate objective should not only produce useful exploration, but ideally reproduce measurable properties of animal behavior and neural activity \citep{keller2025intrinsic}.

Recent whole-brain recordings in \textit{Drosophila} provide a particularly useful ``ground-truth'' target for these kinds of comparisons. \citet{brezovec2023volitional} recorded neural activity while head-fixed flies explored spontaneously on a spherical treadmill in darkness. They found that interhemispheric activity in a subregion of the Inferior Posterior Slope (IPS) was weakly but significantly predictive of future left-versus-right turn direction approximately 15 to 8 seconds \emph{before} a turn. This activity emerged well before the onset of widespread locomotion-related activity across the brain. This provides two complementary constraints on computational hypotheses of internally generated behavior; namely, the spatial and temporal dynamics of the agent's behavior and internal representations of \emph{future} actions.

Methods in computational ethology that describe spontaneous behavior of recurring behavioral syllables~\citep{wiltschko2015mapping,mathis2018deeplabcut,hsu2021b,pereira2022sleap} motivate us to consider the level of behavioral abstraction at which the intrinsic-drive methods should operate.
These syllables exhibit organizational structure across multiple timescales, motivating our implementation of a \emph{hierarchical} policy distribution over motor primitives. These motor primitives are then implemented at the joint-torque level by a low-level central pattern generator (CPG)~\citep{mantziaris2020cpg, chun2021tripod, LobatoRios2022} and kinematic replays of real fly behaviors, such as grooming (Fig. \ref{fig:fig_2_overview}b). 
This separates exploratory action selection from the mechanics of locomotion and allows intrinsic objectives to operate over behaviorally meaningful variables rather than individual motor commands.

We extend the goal-driven modeling approach of computational neuroscience \citep{yamins2016using, richards2019deep, feather2025neuroaituring}, in which candidate objectives are instantiated in artificial networks and compared against neural recordings, to intrinsic motivation \citep{keller2025intrinsic}. We require that a candidate objective account for behavior as well as neural activity. Our agents are embodied in NeuroMechFly v2, a biomechanical model of \textit{Drosophila} simulated in FlyGym, with vision, olfaction, proprioception, and articulated body dynamics \citep{LobatoRios2022, WangChen2024} (Fig.~\ref{fig:fig_1_overview}). We train recurrent policies with $20$ intrinsic drive methods (IDM) spanning curiosity, entropy, and competency objectives (Fig.~\ref{fig:fig_2_overview}a). All methods share the same embodiment, architecture, training environment, and training budget. We then freeze each policy and evaluate it in a virtual tethered-ball assay that reproduces the biological experiment of \citet{brezovec2023volitional} (Fig.~\ref{fig:fig_1_overview}b). For each agent, we compare turning behavior with that of the flies, test whether internal representations predict future turn direction on the timescale of the IPS signal, compare this with what ongoing behavior alone predicts, ask whether its low-dimensional activity shares IPS's sum-difference relationship to angular and forward velocity.

\paragraph{Contributions} To summarize, our contributions are as follows:
\begin{itemize}
    \item \textbf{We implement hierarchical, intrinsically motivated agents embodied in a biomechanically-realistic fly body, in which $\mathbf{20}$ intrinsic objectives operate over seven ethologically motivated behavioral primitives.} 
    \item \textbf{We show that representations of future action can emerge from intrinsic motivation alone, without an externally specified task or reward.} In particular, several intrinsically motivated agents develop internal states that predict turn direction seconds in advance, paralleling the prospective signal observed in the fly IPS.
    \item \textbf{Among the intrinsic objectives tested, unsupervised skill discovery provides the strongest computational account of spontaneous \emph{population-wide} fly behavior and its neural dynamics.} DIAYN, hierarchically operating on behavioral primitives, jointly reproduces future-action predictivity, the relationship between neural and behavioral predictability, inter-animal variability in turning behavior, and low-dimensional activity structure related to forward and angular velocity.  
    \item \textbf{We demonstrate that behavioral predictability and average locomotor statistics alone are insufficient to distinguish candidate explanations of spontaneous action.} Objectives that generate predictable behavior or match mean turning bias can fail to reproduce the fly’s prospective internal signal; jointly evaluating behavioral variability and internal future-action predictivity distinguishes DIAYN from these alternatives. 
\end{itemize}

\section{Background and Related Work}

\subsection{Intrinsically generated and hierarchically structured animal behavior}
Animals generate spontaneous behavior in the absence of explicit external reward or immediate instructive cues. Such cases provide a valuable setting for studying internally generated action selection. 
The field of computational ethology has shown that the behavior of animals such as the fly~\citep{berman2014mapping, berman2016predictability, katsov2017dynamic}, mouse~\citep{wiltschko2015mapping}, zebrafish~\citep{marques2018zebrafish} and \emph{C. elegans}~\citep{kaplan2020nested} can be decomposed into \textit{behavioral syllables} ~\citep{anderson2014computationalethology}. 
A behavioral syllable is simply a recurring, stereotyped pattern of movement that can be reused and combined with other syllables to form more complex behavior. Recent studies have shown behavior exhibits hierarchical organization across syllables and across timescales. 
Specifically, across species, animal behavior can be represented at multiple nested levels of abstraction, with fine grained behavioral states carrying predictive information over short timescales and coarser groupings being informative over longer timescales~\citep{berman2016predictability, kaplan2020nested,mathis2018deeplabcut,hsu2021b,pereira2022sleap}. 
Advances in whole-brain neuroimaging of awake and behaving animals has provided unprecendented access to the neural dynamics of complex behavior  ~\citep{mu2019glia, mann2021coupling, aimon2023global, brezovec2024mapping}. For example, \citet{mu2019glia} image over 200,000 cells spanning neurons and astrocytes in autonomously behaving larval zebrafish, finding that glial cells play a crucial role in behavior by accumulating sensorimotor prediction error signaled by neurons. Here, we utilize recordings in flies during autonomous behavior by \citet{brezovec2023volitional} showing that across bilaterally matched, locomotion-related superclusters distributed around the brain, summed bilateral activity was highly correlated with forward velocity, whereas interhemispheric activity differences were highly correlated with angular velocity. Crucially, this asymmetry emerges much earlier in a subregion of the IPS than the rest of the brain, around 15 to 8 seconds before a turn. Using this region's activity, future turn direction can be predicted above chance. These findings provide measurable behavioral and representational signatures of internally generated action against which computational models can be evaluated. The multiscale organization of animal behavior motivates us to test a hierarchical architecture in which intrinsic drive operates over higher-level locomotor primitives while fixed low-level controllers execute joint dynamics. 

\subsection{Computational theories of intrinsic motivation}

While the complete tapestry of intrinsic motivations that drive autonomous behavior in animals is not yet understood, multiple computational theories have been proposed to capture some essential qualities of exploration. We organize the methods by the primary quantity they reward into three broad families: curiosity (latent prediction-error), entropy (replay diversity), and competency (policy identifiability). This taxonomy is intended as an organizational device rather than a strict partition of mutually exclusive principles. 

\paragraph{Curiosity} Animals are driven by a desire to understand their environment and their place within it. This can be described as \textit{information-seeking} behavior, which can be formalized in a variety of ways \citep{kidd2015psychology}. Most intuitively, an animal can learn a forward model of sensory or sensorimotor dynamics and use its prediction-error or uncertainty over time to quantify how much it "understands" things and makes progress thereafter \citep{schmidhuber2009driven}. In reinforcement learning, this idea was taken seriously and developed early on by Schmidhuber \citep{schmidhuber1991adaptive, schmidhuber1991curiosity, schmidhuber1997interesting, schmidhuber1999artificial}, leading to the development of numerous variants and adaptations. In this paper, we choose a representative mix of curiosity methods, from prediction-error to information-gain to learning progress. This includes \textit{Disagreement} \citep{pathak2019disag}, \textit{$\gamma$-Progress} \citep{kim2020gammaprogress}, \textit{Intrinsic Curiosity Module} (ICM) \citep{pathak2017icm}, \textit{Model-based Active Exploration} (MAX) \citep{shyam2019max}, \textit{Never Give Up} (NGU) \citep{badia2020ngu}, \textit{Random Network Distillation} (RND) \citep{burda2019rnd}, and \textit{Curiosity-driven
Exploration via Temporal Contrastive Learning} (C-TeC) \citep{mohamed2026ctec}. The essential object in all algorithms is some predictive model, and they differ in what exactly is predicted or how that model is used to form the intrinsic reward. 
\begin{figure}[t!]
  \centering
  \includegraphics[width=\textwidth, page=1]{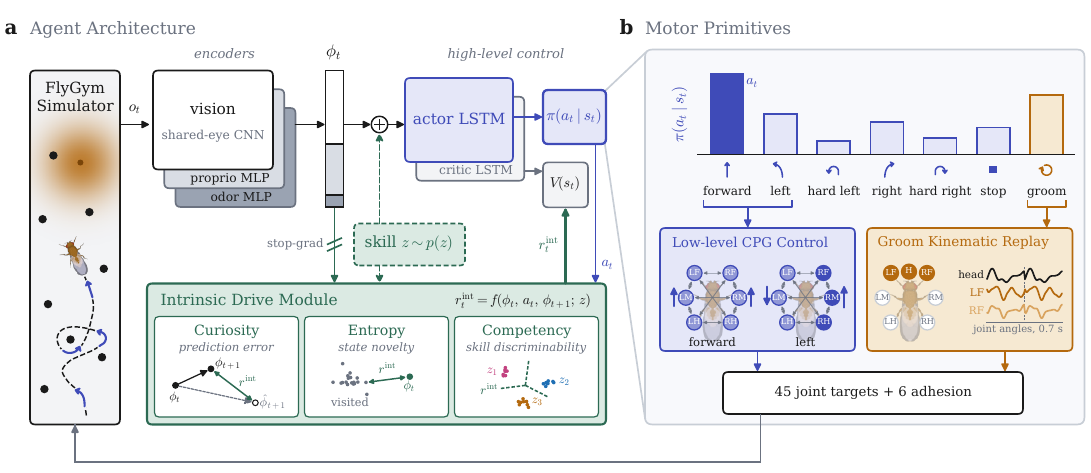}
  \caption{\textbf{Agent Architecture Overview.}
\textbf{(a)} The FlyGym simulator provides visual, proprioceptive, and olfactory observations from a biomechanically realistic fly body. Modality-specific CNN and MLP encoders embed these observations into a shared representation $\phi_t$, which is provided to the actor--critic and intrinsic drive module. When applicable, the policy is additionally conditioned on a sampled skill $z$. The intrinsic drive module computes an intrinsic reward from the encoded state transition and selected action. The actor outputs a categorical distribution over discrete motor primitives, while the critic estimates the state value. The policy selects a new motor primitive every 100\,ms.
\textbf{(b)} Locomotor primitives are translated into joint-level commands by a low-level central pattern generator (CPG) controller, while grooming is executed using a kinematic replay. These motor commands are applied within the FlyGym physics simulation, which is integrated with a timestep of 100\,$\mu$s. Thus, each policy action is executed for 1,000 physics steps before the next action is selected.}
  \label{fig:fig_2_overview}
\end{figure}
\paragraph{Entropy} Not all kinds of exploration are best explained by a drive to understand or predict the world. On the contrary, pursuing sensory novelty for novelty's sake means that one might not care why or how a stimulus was produced. For example, mice enjoy highly variable sensory stimuli independently of learning. This preference persists, and even increases, after familiarity with the stimulus contingencies \citep{olsen2012stimulus, dickson2020stimulus}, suggesting animals are driven by the overall entropy of experience. Thus, entropy-based methods reward novelty with respect to the state or feature distribution, encouraging broad exploration coverage independent of preditability under an internal model. This family includes \textit{Active Pretraining} (APT) \citep{liu2021apt}, \textit{Reinforcement Learning with Prototypical Representations (Proto-RL)} \citep{yarats2021protorl}, and \textit{State Entropy Maximization with Random Encoders for Efficient Exploration}
(RE3) \citep{seo2021re3}.

\paragraph{Competency} Animals often behave to specifically isolate the role of their actions in generating sensory observations, akin to learning a causal model of the world in which controllability plays a central role. Competency aims to capture this drive to interact with the environment effectively and masterfully \citep{white1959motivation}. This was coined as \textit{Empowerment} and formalized by \citep{klyubin2008emppot} as the channel capacity between actions and sensory states, encouraging the agent to occupy states in which its actions have influence on future states \citep{mohamed2015empgainfwd, zhao2021empest}. We consider some additional variants, such as \textit{Empowerment Progress} \citep{beckerehmck2021empgain} and \textit{Process Empowerment} \citep{tiomkin2025empproc}, which track the channel capacity over time or maximize the capacity under the policy distribution, respectively. We evaluate forward, gain, potential, and process variants of empowerment. Skill-discovery (or options-discovery) methods apply the idea of empowerment to entire policies rather than individual actions by introducing a hierarchical skill variable that indexes different policies. Then, mutual information is maximized between each skill and the states induced under that skill. This family includes \textit{Contrastive Successor Features} (CSF) \citep{zheng2025csf}, \textit{Dynamics-Aware Unsupervised Discovery of Skills} (DADS) \citep{sharma2020dads}, Diversity is All You Need (DIAYN) \citep{eysenbach2019diayn}, \textit{Variational Autoencoding Learning of Options by Reinforcement} (VALOR) \citep{achiam2018valor}, and \textit{Variational Intrinsic Control} (VIC) \citep{gregor2017vic}.

\subsection{Artificial neural networks as hypotheses of biological computation}

A major approach in computational neuroscience and NeuroAI is to treat task-optimized artificial neural networks as candidate models of biological computation \citep{yamins2016using,richards2019deep}. 
Representations learned by ANNs have been compared to single brain areas, such as the primate visual cortex \citep{yamins2013hierarchical,khaligh2014deep,cadena2019deep}, mouse visual cortex \citep{bakhtiari2021functional,nayebi2023mouse}, the rodent entorhinal system \citep{banino2018vector,nayebi2021entorhinal}, auditory cortex \citep{kell2018task,feather2019metamers} and the motor system \citep{sussillo2015neural,michaels2020goal}.
More recently, NeuroAI has leveraged embodied AI to bridge computational neuroscience with behavior, including virtual animal models such as the rodent~\citep{merel2019deepneuroethologyvirtualrodent,aldarondo2024virtual}, the fruit fly~\citep{vaxenburg2025whole,LobatoRios2022, WangChen2024}, and the zebrafish~\citep{keller2025intrinsic}.
Following \citet{keller2025intrinsic}, our agents are trained entirely through a self-supervised objective (intrinsic motivation) unlike imitation-based virtual animals, thereby providing a \emph{normative} account of autonomous behavior by replicating neural and behavioral dynamics at test time. By implementing our intrinsically motivated agents hierarchically over a range of the fly's behavioral primitives, we extend the approach introduced by \citet{keller2025intrinsic} beyond two coarse behavioral states (activity and passivity) to a much larger repertoire. 

\begin{figure}[t]
  \centering
  \includegraphics[width=\textwidth]{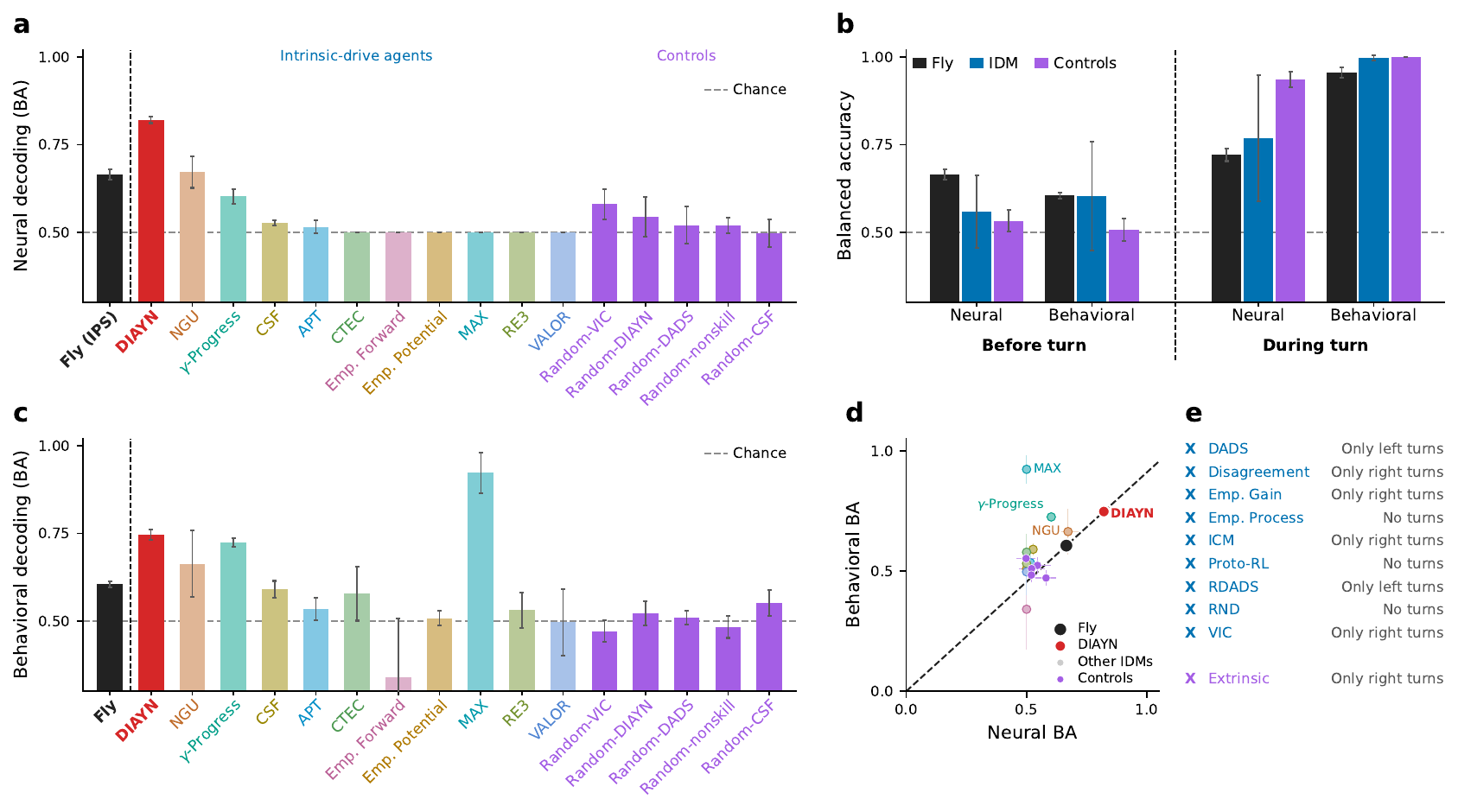}

\caption{
  \textbf{Turn-direction decoding from neural and behavioral data.}
  Prospective coding is quantified as balanced accuracy in predicting future left/right turns.
  \textbf{(a,c)} Neural (a) and behavioral (c) prospective coding in the fly and models using elastic-net logistic regression in the \beforeturn{} window.
  \textbf{(b)} Neural and behavioral decoding before (\beforeturn{}) and during (\duringturn{}) turns.
  \textbf{(d)} Behavioral versus neural prospective coding. The dashed line defines the fly's behavioral-to-neural accuracy ratio, which DIAYN closely matches.
  \textbf{(e)} Models with no turns or only one turn direction, precluding stratified cross-validation.
}

  \label{fig:decoding_results}
\end{figure}

\section{Methods}

\subsection{Hierarchical Agent Architecture and Biomechanical Model}

Our agents are embodied in NeuroMechFly v2, a neuromechanical digital twin of an adult female \textit{Drosophila} simulated in FlyGym with MuJoCo at a timestep of $10^{-4}$ seconds \citep{LobatoRios2022, WangChen2024, wang2026embodied, todorov2012mujoco}.
The policy receives compound-eye vision from pale- and yellow-type ommatidia, a bilateral olfactory signal, and a proprioceptive state of joint angles, joint velocities, and body kinematics (Fig.~\ref{fig:fig_1_overview}c--e; App.~\ref{app:obs}).
Control is divided hierarchically (Fig.~\ref{fig:fig_2_overview}). Every $100$\,ms, the policy selects one of seven behavioral primitives: forward, left, hard left, right, hard right, stop, and groom. The six locomotor primitives set left and right descending signals that modulate a central pattern generator (CPG), whose phase indexes step trajectories recorded from real walking flies~\citep{mantziaris2020cpg, chun2021tripod}. The groom primitive executes a kinematic replay of recorded fly grooming (App.~\ref{app:control}). Behavior therefore arises from the policy chaining primitives across many decisions, while joint-level mechanics are fixed across all intrinsic objectives.
Modality-specific encoders produce a $310$-dimensional feature vector that feeds separate actor and critic LSTMs ($256$ units each), followed by MLP heads. The actor outputs a categorical distribution over the seven primitives (Fig.~\ref{fig:fig_2_overview}a; Table~\ref{tab:policy_params}). For skill-conditioned methods, the skill vector is appended to the input of both LSTMs but withheld from the intrinsic reward module.

\begin{figure}[t]
  \centering
  \includegraphics[width=\textwidth, page=1]{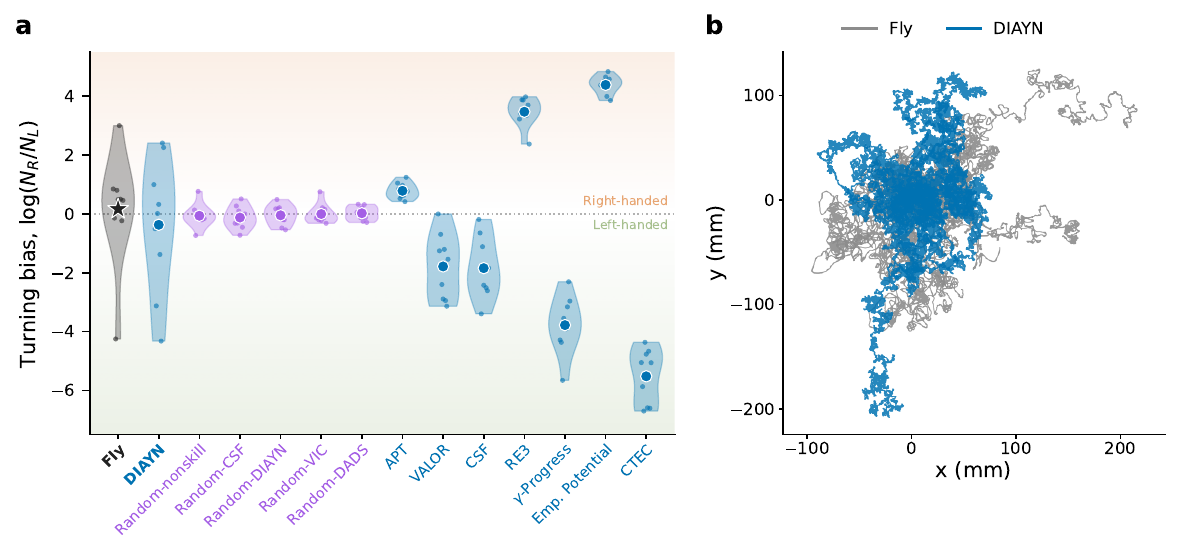}
  \caption{ \textbf{Turning Bias.} \textbf{(a)} Distribution of turning bias across evaluation rollouts for each method. Points indicate individual rollouts, with the mean and range shown for each method. DIAYN produces a broad distribution of turning biases spanning more of the fly-to-fly variability than the other evaluated methods. Methods with a rollout containing no turns or turns in only one direction are excluded because $\log(N_R/N_L)$ is undefined for those rollouts. \textbf{(b)} Virtual x-y trajectories integrated from the ball state of the nine flies and the nine DIAYN evaluation skills drawn from a common origin.}
  \label{fig:bias_and_path}
\end{figure}

\subsection{Intrinsic Drive Methods and Training}

We train $20$ intrinsically motivated policies using methods across three broad families: curiosity, entropy, and competency (Sec.~2.2; App.~A.4). The intrinsic reward is the only learning signal the policy receives. We use Recurrent PPO \citep{raffin2021stablebaselines3, schulman2017ppo} in $2{,}048$ parallel environments, procedurally generated arenas with three odor sources and $12$ to $50$ obstacles which vary the visual, olfactory, and proprioceptive input per episode (Fig.~\ref{fig:fig_1_overview}a). All methods share the same architecture, environment, hyperparameters, and budget of $50$ million policy decisions (Table~\ref{tab:ppo_params}), and all analyses evaluate the final checkpoint. As controls, we evaluate randomly initialized policies with the same architectures as well as a policy trained on an extrinsic odor navigation task (App.~\ref{app:controls}).

Because DIAYN provides the strongest joint account of the behavioral and neural signatures examined below, we describe its objective here to clarify how skill discovery interacts with our hierarchical motor architecture; implementation details for DIAYN and all other intrinsic-drive methods are provided in App.~\ref{app:idms}.

\paragraph{DIAYN and Hierarchical Skill Discovery}
DIAYN \citep{eysenbach2019diayn} learns a skill-conditioned policy by maximizing the objective 
\begin{equation}
\begin{aligned}
r(\theta)
&= I(S; Z) + H(A \mid S, Z)\\
&= H(Z) - H(Z \mid S) + H(A \mid S,Z),
\end{aligned}
\label{eq:diayn}
\end{equation}
where \(Z\) denotes the latent skill (drawn from a uniform categorical distribution), \(S\equiv \phi\) the agent sensory-state (Fig. \ref{fig:fig_2_overview}), and \(A\) the action at time $t$. In our hierarchical architecture, \(A\) corresponds to one of the seven behavioral motor primitives rather than a joint-level motor command. Here, \(I(\cdot;\cdot)\) denotes mutual information, \(H(\cdot)\) entropy, and \(H(\cdot\mid\cdot)\) conditional entropy. The first term, \(H(Z)\), encourages a diverse skill prior and is constant in our implementation because skills are sampled uniformly. The second term, \(-H(Z\mid S)\), encourages different skills to induce distinguishable states, such that the skill can be inferred from the states it visits. The final term, \(H(A\mid S,Z)\), encourages high-entropy action selection within each skill. Thus, each skill learns a distinct strategy for selecting among the same shared set of motor primitives as a function of the sensory state. Equation \ref{eq:diayn} is optimized in expectation by PPO using the per-step intrinsic reward \(r_t=\log q_\theta(z\mid\phi_t)-\log \pi(a_t\mid \phi_t, z)\), with $q_\theta$ the learned skill discriminator within the IDM and $\pi$ the high-level policy. 

\subsection{Tethered-Ball Evaluation}

Frozen policies are evaluated in a virtual version of the assay of \citet{brezovec2023volitional}, where the fly is head-fixed on a $9$\,mm air-supported ball, and angular and forward velocity are read from the ball simulator state (Fig.~\ref{fig:fig_1_overview}b). To match the darkness and absence of odor in the biological experiment, visual and olfactory inputs are set to zero. For each policy we collect nine $30$-minute rollouts, matching the nine flies and $270$ minutes of biological data. All policies are evaluated via sampling. Skills are sampled once at the start of each episode and resampled whenever the number of decisions reaches the method's skill horizon. The LSTM hidden state is not reset during the evaluation. Details are given in App.~\ref{app:eval}.

\section{Results}

Behavioral data is processed identically for agents and flies matching the procedure of \citet{brezovec2023volitional}. First, velocities are smoothed with a $500$\,ms Savitzky-Golay filter \citep{savitzky1964smoothing}. Then, turns are detected as peaks in absolute angular velocity above $200^\circ/\mathrm{s}$ \citep{brezovec2023volitional}. Each evaluation rollout or fly experiment is summarized by its turning bias, $\log(N_R/N_L)$. To test for a prototypical planning representation, we decode future turn direction from the activity in the \beforeturn{} window using elastic-net penalized logistic regression \citep{zou2005elasticnet,pedregosa2011scikit} with five-fold stratified cross-validation and report balanced accuracy \citep{brodersen2010balanced}.
Decoder hyperparameters are tuned once on the fly data and fixed for all agents. For agents, we evaluate seven candidate internal representations: the actor LSTM, critic LSTM, policy MLP, and value MLP representations individually, together with three concatenations of these recorded features. The largest concatenation additionally includes the learned observation embedding. As a separate input-level control, we also decode from the raw proprioceptive observation, with vision and odor zeroed. For the biological fly, performance is decoded from IPS activity (App.~\ref{app:bio}).
Because future turns may be predictable from current behavior alone, we fit the same decoder to mean angular and forward velocity in the same window to identify policies that converged to a relatively stable, predictable behavioral strategy  (App.~\ref{app:behavior_analysis}).
Finally, to compare the representation geometry without assuming an anatomical correspondence, we apply PCA to each agent's activity in its highest-performing neural layer for decoding. We test whether pairs of components relate to angular and forward velocity as the IPS hemispheres do (App.~\ref{app:pca}). 
\paragraph{Neural and Behavioral Decoding} Using the network activations from the models in the tethered-ball evaluation, we follow the same conceptual test used for the biological fly, asking whether internal activity before a turn predicts its eventual direction and plot the prospective coding in Fig. \ref{fig:decoding_results}a. The biological fly IPS has a neural prospective coding of $66.5\pm1.4\%$ while DIAYN and NGU of the IDM-trained models approach or exceed this performance with $82.1\pm0.9\%$ and $65.4\pm1.8\%$ respectively (Fig. \ref{fig:decoding_results}a). Many IDM-trained policies are near $50\%$ prospective coding or had an undefined prospective coding due to never having a detected turn or only having detected turns of a single direction. Neither condition permits stratified fold cross-validation for the accuracy calculation.
Analogously, we decode turn direction from behavior. Using the mean angular ($\omega$) and forward ($v$) velocity of the ball in the before-turn window, we fit a logistic regression classifier to decode the future direction of the turn and plot the prospective coding in Fig. \ref{fig:decoding_results}c. For the flies, behavior alone predicts future turn direction with a prospective coding of $60.5\pm0.9\%$. For the IDM-trained policies, many exhibit behavior that is predictive of future turn direction and decodable including MAX ($92.2\pm5.9\%$), DIAYN ($74.7\pm1.4\%$), $\gamma$-progress ($72.4\pm1.2\%$) and NGU ($66.3\pm9.4\%$). Nine of the $20$ methods produce no turns or turns in only one direction during evaluation (Fig.~\ref{fig:decoding_results}e), itself a behavioral mismatch with the fly.
\begin{figure}[t]
  \centering
  \includegraphics[width=\textwidth, page=1]{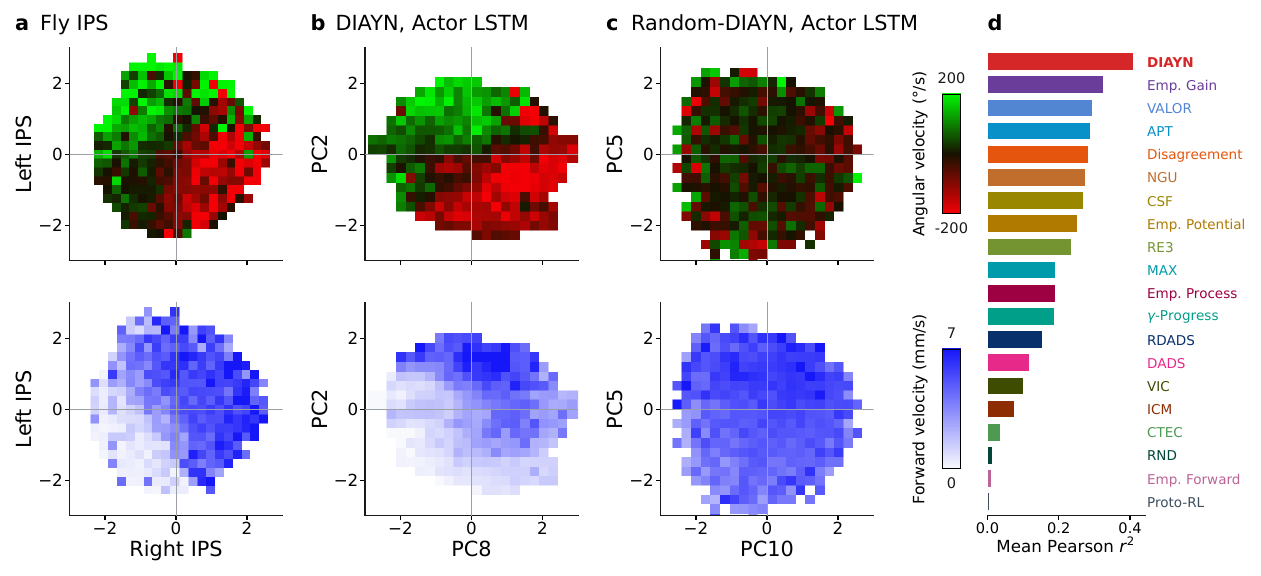}
  \caption{\textbf{Low-dimensional internal structures relate to $\omega$ and $v$}. 
  \textbf{(a)} Reconstructed from \citet{brezovec2023volitional}, the IPS of the fly brain has an interhemispheric relationship of z-scored activity that is correlated with angular (interhemispheric difference) and forward velocity (interhemispheric sum). 
  \textbf{(b)} As a functional analogue of the bilateral supercluster pairs in \citet{brezovec2023volitional}, we use PCA to find principal components in the network activation of the LSTM layer of the Actor in the policy whose difference correlates with angular velocity and whose sum correlates with forward velocity. This structural relationship appears in DIAYN's Actor LSTM layer \textbf{(b)}  but, does not appear in 
  \textbf{(c)} the randomly initialized DIAYN model. \textbf{(d)} We rank each IDM by the mean of the squared Pearson correlations of its angular and forward heatmaps with the fly's. Higher is a closer match.}
  \label{fig:rep_analysis}
\end{figure}
\paragraph{Performance Ratio} These results distinguish IDMs that produce internally predictive representations from those whose behavior is simply easy to predict from past behavior. Critically, an agent whose behavior resembles the fly's should show a similar ratio of neural to behavioral decoding accuracy: internal states should carry predictive information beyond behavior, while behavior itself is structured but not trivially predictable. Therefore, Fig. \ref{fig:decoding_results}d considers neural and behavioral predictability jointly. Among the trained agents, DIAYN and NGU are the only IDMs that match or exceed the neural and behavioral predictability of the fly. Unlike NGU, DIAYN also maintains the same ratio of predictability as the fly. 

\paragraph{Turning Bias} Models that match the mean statistics of the flies do not match the flies' population variability. Fig. \ref{fig:bias_and_path}a shows each model's distribution of turning biases. Models whose mean bias is close to the flies', such as APT and the random controls, do not reproduce the fly population of turning biases. In contrast, DIAYN's skills span a much wider range of turning biases, yielding a more similar state space coverage to the flies (Fig.~\ref{fig:bias_and_path}b) based on trajectory coverage. DIAYN holds one skill per $30$-minute rollout, so its nine rollouts sample nine skills, paralleling the nine flies. CSF and Random-DIAYN use the same schedule yet show far narrower bias distributions (Fig.~\ref{fig:bias_and_path}a), indicating that the fly-like spread arises from the learned objective rather than the sampling scheme.

\paragraph{Internal Representations} The internal activation dynamics of the models were also compared with the fly neural dynamics. As shown in Fig. \ref{fig:rep_analysis}a, neural activity in the fly IPS forms separable structures whose difference and sum is correlated with angular and forward velocity, respectively. Fig. \ref{fig:rep_analysis}b shows that trained DIAYN recovers a similar relationship, whereas this organization is absent in the random-initialized control (Fig. \ref{fig:rep_analysis}c). For trained DIAYN, the difference of the selected component pair correlates with $\omega$ ($r = 0.153$) and their sum with $v$ ($r = 0.413$), comparable to the fly IPS ($r = 0.191$ and $0.429$). In Random-DIAYN, the best pair reaches only $r = 0.021$ and $0.113$.

\section{Discussion}
The central implication of our results is that behavioral diversity and prospective internal activity may reflect a shared computational objective. Methods that match average turning bias or generate highly predictable behavior need not reproduce the accompanying predictive neural representation (Figs.~\ref{fig:decoding_results} and~\ref{fig:bias_and_path}). DIAYN's joint correspondence with these properties therefore motivates an explanation of autonomy that addresses both the diversity of spontaneous behavior and the internal organization that makes future actions predictable.

In our hierarchical implementation, DIAYN learns strategies for selecting among seven motor primitives, so its objective concerns the organization of behavior above the level of individual motor commands. It rewards skills that induce distinguishable sensory states while encouraging high-entropy action selection within each skill. Because each skill is held fixed throughout a training or evaluation episode, we interpret it as a persistent exploratory strategy consistent with the locomotor handedness observed in individual flies \citep{Buchanan2015Handedness, Ayroles2015Idiosyncrasy, deBivort2022BehavioralIndividuality}. \citet{brezovec2023volitional} show wide individual variation in fly exploration strategies, as reflected in their turning biases, and remarkably, DIAYN's learned repertoire of skills captures this variation quite well (Fig.~\ref{fig:bias_and_path}). The repertoire thus offers a model of behavioral diversity across a \emph{population} of flies. One possible explanation is that the fly brain's limited computational capacity favors a single, relatively fixed exploratory strategy per individual, while the diversity of strategies needed for survival is realized through their distribution across individuals to \emph{maximize} state-space coverage as a \emph{population}. Under this hypothesis, evolutionary pressures favor diverse individual strategies that collectively maximize coverage of sensory space. Handedness and other persistent behavioral tendencies we observe in individuals could also be interpreted as individual expressions of this population-level diversity.

Skill discovery may also help explain why internal activity becomes predictive of future turns. Although DIAYN's intrinsic reward is evaluated at individual time steps, PPO optimizes the policy to maximize cumulative intrinsic reward over trajectories. This optimization can favor sustained patterns of behavior that keep the sensory states induced by different skills distinguishable over time. In our hierarchical controller, such patterns can arise through persistent, skill-dependent tendencies to select particular motor primitives. The recurrent state may therefore encode the ongoing exploratory strategy and carry information about turns that occur seconds later. Under this hypothesis, prospective activity reflects the ongoing organization of a behavioral strategy rather than moment-by-momemnt movements, consistent with the findings in real fruit flies in \citet{brezovec2023volitional}. Hierarchical skill-discovery therefore provides a concrete computational framework in which animal-like behavioral diversity and future-action representations emerge together under a common intrinsic objective.

\section{Conclusion}
We introduce a framework for reverse-engineering animal autonomy that combines a neuromechanical digital twin, hierarchical control over behavioral primitives, and joint comparison with fly behavior and neural recordings. Across $20$ intrinsic objectives, DIAYN provides the strongest account of individual differences in turn decisions, prospective neural activity, and interhemispheric sum-difference encoding of kinematics. Importantly, these properties emerge entirely from optimizing a self-supervised, intrinsic goal without training on external rewards or fly data. Our results show that individual behavioral variability and shared representations of future action can arise from the same computational objective. DIAYN links these findings through persistent exploration strategies whose internal states predict future turns seconds in advance. This suggests a novel perspective on the role of individual differences in optimizing exploration at the population level, an efficient strategy for survival of a species when the capacity of individual members is limited. These results establish the importance of using both neural and behavioral data in distinguishing computational models, the role of hierarchy in designing agents that naturally express complex behavior, and the potential of embodied agents and intrinsic reinforcement learning to understand principles of animal autonomy. 

\subsection*{Limitations and Future Work}
One main limitation of our approach is the choice to constrain the space of behaviors available to each policy to the predefined locomotor primitives and low-level controllers. While this reflects the hierarchical organization of animal behavior, it limits the learned behavior primarily to 2-D locomotion. Future work could learn the primitives jointly with their high-level selection, allowing the IDM to shape its low-level repertoire, or expanding the primitives to allow flight, potentially extending the pre-existing implementation of \citet{vaxenburg2025whole}. 

Another opportunity is to extend our framework to account for the whole-brain dynamics of locomotion \citep{brezovec2024mapping}. Due to our focus on IPS, a simple training environment was sufficient to reproduce key features of those neural responses. 
A \emph{whole-brain} pursuit would force us to confront an ethnologically relevant environment with variable terrain, opportunities for manipulation and resources that may favor different intrinsic objectives and thereby generate a broader repertoire of behaviors---which is possible to explore within the agent-based framework we build here.

\subsection*{AI use statement}
In this work, we used generative AI tools to assist with research ideation and execution, including implementation of aspects of the project. Additionally, we used generative AI tools to aid and polish writing, retrieve and discover related work, and create and format plots. AI-assisted implementation work was independently reviewed by the primary authors. We take responsibility for the final content of this work, including text, claims or artifacts produced with the aid of generative AI.

\subsubsection*{Author Contributions}
D.K. implemented the training architecture, environments, and policy on top of FlyGym. R.K. implemented the IDMs. D.K. and A.K. implemented the ball evaluation environment. F.P. provided preprocessing and data utilities for the fly neural data. A.N., R.K. and D.K. designed the experiments and analysis. A.N. originated and supervised the project. 

\subsubsection*{Acknowledgments}
We thank Bella Brezovec, Gert-Jan Both, Rob Johnson, Srini Turaga, and Roman Vaxenburg for helpful discussions in the early phases of this project. We thank Sibo Wang-Chen for helpful discussions and technical guidance on adapting locomotion controllers to the FlyGym v2 API.
D.K. was supported with funding and resources from Northrop Grumman, the United States Department of Defense High Performance Computing Modernization Program (HPCMP), and the State of Maryland. R.K. was supported in part by the Simons Collaboration in Ecological Neuroscience (SFI-AN-NC-SCN-00007276) and in part from funds provided by the National Science Foundation and by DoD OUSD (R \& E) under Cooperative Agreement PHY-2229929 (The NSF AI Institute for Artificial and Natural Intelligence, ARNI). 
A.N. was supported in part by the Burroughs Wellcome Fund (CASI award), Foresight Institute, and Protocol Labs.

\bibliography{iclr2027_conference}
\bibliographystyle{iclr2027_conference}

\appendix
\appendix

\section{Additional Experimental Details}
\label{app:details}

\subsection{Sensory Observations}
\label{app:obs}

\paragraph{Proprioception.} The proprioceptive observation is
\begin{equation}
\mathbf{o}^{\mathrm{prop}}_t = \left[\mathbf{q},\ \dot{\mathbf{q}},\ h,\ \mathbf{g}_{\mathrm{body}},\ \mathbf{v}_{\mathrm{body}},\ \boldsymbol{\omega}_{\mathrm{body}}\right] \in \mathbb{R}^{100},
\end{equation}
where $\mathbf{q}, \dot{\mathbf{q}} \in \mathbb{R}^{45}$ are the angles (rad) and angular velocities (rad/s) of the 45 position-actuated degrees of freedom, $h$ is the height of the body root (mm), $\mathbf{g}_{\mathrm{body}} \in \mathbb{R}^3$ is gravity in the body frame, and $\mathbf{v}_{\mathrm{body}}, \boldsymbol{\omega}_{\mathrm{body}} \in \mathbb{R}^3$ are the linear (mm/s) and angular (rad/s) velocities of the body root. The 45 degrees of freedom are seven per leg (thorax--coxa yaw, pitch, and roll; coxa--trochanter pitch and roll; femur--tibia pitch; tibia--tarsus pitch), ordered LF, LM, LH, RF, RM, RH, followed by neck pitch, roll, and yaw (Fig.~\ref{fig:fig_1_overview}d).

\paragraph{Vision.} Each compound eye is rendered by a fisheye camera ($512\times450$ pixels, $157^\circ$ field of view) and read out as 721 ommatidia, 505 yellow-type and 216 pale-type \citep{LobatoRios2022, WangChen2024}. Each ommatidium takes the mean of the green (yellow-type) or blue (pale-type) channel over its pixels, scaled to $[0,1]$. For computational efficiency, the ommatidial map is resampled from $512\times450$ to a $32\times32$ grid by nearest-neighbor sampling at each grid-pixel center; grid pixels outside the lattice are zero, and 623 of the 721 ommatidia are represented. The visual input is a $2\times2\times32\times32$ tensor (eye $\times$ ommatidium type $\times$ height $\times$ width; Fig.~\ref{fig:fig_1_overview}e).

\paragraph{Olfaction.} Odor intensity at position $\mathbf{x}$ is $I(\mathbf{x}) = \sum_k A_k \lVert \mathbf{x} - \mathbf{x}_k \rVert^{-2}$, summed over sources $k$ with peak amplitude $A_k = 1$. Intensity is sampled at the left and right maxillary palps and antennae and combined on each side as $0.1\,I_{\mathrm{palp}} + 0.9\,I_{\mathrm{antenna}}$, giving a two-dimensional left/right observation (Fig.~\ref{fig:fig_1_overview}c).

\paragraph{Normalization.} Odor and proprioceptive observations are normalized with running means and variances and clipped at $\pm10$. Visual input is not normalized.

\subsection{Hierarchical Locomotor Control}
\label{app:control}

Every $0.1$\,s, the policy selects one of seven behavioral primitives (Table~\ref{tab:primitives}; Fig.~\ref{fig:fig_2_overview}b). Six set a pair of descending signals $(d_L, d_R)$, applied to the three left and three right legs respectively; the seventh triggers a grooming replay.

\begin{table}[t]
\centering
\small
\begin{tabular}{rlrrl}
\toprule
Index & Primitive & $d_L$ & $d_R$ & Controller \\
\midrule
0 & Left       & $-1.0$ & $1.2$  & CPG turning \\
1 & Forward    & $1.0$  & $1.0$  & CPG walking \\
2 & Right      & $1.2$  & $-1.0$ & CPG turning \\
3 & Stop       & $0$    & $0$    & CPG, amplitude decays to zero \\
4 & Groom      & $0$    & $0$    & Joint replay overrides CPG \\
5 & Hard left  & $-1.2$ & $1.2$  & CPG turning \\
6 & Hard right & $1.2$  & $-1.2$ & CPG turning \\
\bottomrule
\end{tabular}
\caption{Seven discrete behavioral primitives. The sign of $d$ sets the stepping direction of that side's legs (negative is backward) and $|d|$ sets the stepping amplitude.}
\label{tab:primitives}
\end{table}

\paragraph{Central pattern generator.} Locomotor primitives drive a six-oscillator CPG, a GPU reimplementation of the FlyGym hybrid turning controller \citep{LobatoRios2022, WangChen2024} without its stumbling and retraction correction rules. For leg $i$ with side signal $d_i \in \{d_L, d_R\}$, the phase $\theta_i$ and amplitude $r_i$ evolve as
\begin{align}
\dot{\theta}_i &= 2\pi\nu\, s(d_i) + \sum_j r_j\, w_{ij} \sin\!\left(\theta_j - \theta_i - \varphi_{ij}\right), &
\dot{r}_i &= \alpha\left(|d_i| - r_i\right),
\end{align}
where $s(d) = -1$ if $d < 0$ and $+1$ otherwise, $\nu = 12$\,Hz, $\alpha = 20$, $w_{ij} = 10$ for coupled legs, and the phase biases $\varphi_{ij} \in \{0, \pi\}$ place $\{\mathrm{LF}, \mathrm{LH}, \mathrm{RM}\}$ and $\{\mathrm{LM}, \mathrm{RF}, \mathrm{RH}\}$ in antiphase tripods. The equations are integrated with forward Euler at the physics timestep ($10^{-4}$\,s). Joint targets are
\begin{equation}
\mathbf{q}^{\ast}_i = \bar{\mathbf{q}}_i + r_i\left(\boldsymbol{\psi}_i(\theta_i) - \bar{\mathbf{q}}_i\right),
\end{equation}
where $\boldsymbol{\psi}_i$ is a periodic cubic spline through a single recorded step of untethered fly walking for leg $i$, and $\bar{\mathbf{q}}_i = \boldsymbol{\psi}_i(\pi)$ is the neutral pose. Joints are position-controlled ($k_p = 45$, force range $\pm65$). Leg adhesion (gain 40) is on during stance and off during swing, with the swing interval extended by $\pi/4$. At each reset, phases are drawn from $\mathcal{U}[0, 2\pi)$ and amplitudes are zero. Under the stop primitive, amplitudes decay to zero with time constant $1/\alpha = 50$\,ms while phases continue to advance.

\paragraph{Grooming.} The groom primitive replays a $35$-frame ($0.35$\,s) clip of a recorded, optogenetically evoked grooming bout (aJO-GAL4 $>$ CsChrimson) at $100$\,Hz, driving $17$ degrees of freedom: the three neck joints and the seven joints of each front leg \citep{ozdil2026centralized}.
The CPG receives a stop signal, the middle and hind legs are held at their pose at groom onset with adhesion on for stability, and the body root is held fixed for the duration of the bout. To avoid abrupt movement, the replay blends in linearly from the onset pose over $0.05$\,s; on the next non-groom action, control switches back to the CPG. Grooming is not a macro-action: each groom action lasts one $0.1$\,s decision, consecutive groom actions continue the clip (looping every $35$ frames).

\subsection{Policy Architecture}
\label{app:policy}

Policies use the recurrent actor--critic of sb3-contrib RecurrentPPO \citep{raffin2021stablebaselines3}, with a shared multimodal feature extractor and separate actor and critic LSTMs \citep{hochreiter1997lstm} (Fig.~\ref{fig:fig_2_overview}a; Table~\ref{tab:policy_params}). A CNN  \citep{fukushima1980neocognitron, lecun1998gradient} with weights shared across the two eyes produces a $64$-dimensional embedding per eye, for a $128$-dimensional visual representation. MLPs \citep{rumelhart1986backprop} embed odor ($16$ dimensions) and proprioception (64 dimensions). The odor embedding is concatenated with the raw $2$-dimensional odor observation and the proprioceptive embedding with the raw $100$-dimensional proprioceptive state, yielding modality representations of $128$, $18$, and $164$ dimensions and a $310$-dimensional base feature vector $\phi_t$. This is passed to the actor and critic LSTMs ($256$ hidden units each) and to the intrinsic drive module (IDM). Separate actor and critic MLP heads follow the LSTMs, and the actor outputs a categorical distribution over the seven primitives.

For skill-conditioned methods, the skill $z$ is appended to $\phi_t$ at the input of both LSTMs. The IDM receives only $\phi_t$, never $z$, which prevents the skill discriminator from being solved trivially. Linear and convolutional layers use orthogonal initialization (gain $\sqrt{2}$; $0.01$ for the action head and $1$ for the value head), and the LSTMs use PyTorch's default initialization. The policy without a skill input has $1{,}407{,}576$ parameters; a skill of dimension $d_z$ adds $2{,}048\,d_z$.

\begin{table}[t]
\centering
\small
\begin{tabularx}{\textwidth}{lX}
\toprule
Component & Architecture \\
\midrule
Visual encoder (shared across eyes) & Conv($2\!\to\!32$, $5\times5$, stride 2) -- Conv($32\!\to\!64$, $3\times3$, stride 2) -- Conv($64\!\to\!64$, $3\times3$, stride 2) -- FC($1024\!\to\!64$) \\
Visual embedding & $2 \times 64 = 128$ \\
Olfactory encoder & MLP $2\!\to\!32\!\to\!16$, concatenated with raw input ($18$) \\
Proprioceptive encoder & MLP $100\!\to\!128\!\to\!64$, concatenated with raw input ($164$) \\
Base features $\phi_t$ & $310$ \\
Actor and critic LSTMs & One layer, 256 units each \\
Actor head & $256\!\to\!128\!\to\!128\!\to\!7$, categorical \\
Critic head & $256\!\to\!128\!\to\!128\!\to\!1$ \\
Activations & ReLU (encoders), tanh (heads) \\
Optimizer & Adam, learning rate $3\times10^{-4}$, $\epsilon = 10^{-5}$ \\
\bottomrule
\end{tabularx}
\caption{Policy network architecture.}
\label{tab:policy_params}
\end{table}

\subsection{Intrinsic Drive Methods}
\label{app:idms}

\paragraph{Shared implementation.} All $20$ methods share one pipeline. The IDM receives the base features $\phi_t$, computed by the policy's feature extractor without gradient flow into the extractor, along with the one-hot action $a_t$ and the next features $\phi_{t+1}$. Observations are normalized with statistics frozen at the start of each rollout. Intrinsic rewards are computed online at every decision using the IDM parameters from the previous rollout. After each PPO rollout, the IDM takes one Adam step (learning rate $10^{-4}$) on a minibatch sampled without replacement from that rollout (256 transitions unless stated otherwise), giving $245$ IDM updates over training; VIC and VALOR instead train on completed skill segments, which may span rollouts. The PPO reward is the intrinsic reward alone, with no extrinsic term and no PPO reward normalization; where noted, the intrinsic reward is divided by its running root-mean-square (RMS). Unless stated otherwise, networks use ReLU activations and Xavier-uniform initialization. Table~\ref{tab:idm_summary} summarizes the methods. The descriptions below state each method as implemented and note departures from the original publication where we identified them.

\begin{table}[t]
\centering
\small
\begin{tabular}{llll}
\toprule
Method & Family & Skill prior (resampling) & Trainable IDM parameters \\
\midrule
ICM               & Curiosity   & --                                    & 543,495 \\
RND               & Curiosity   & --                                    & 276,992 \\
Disagreement      & Curiosity   & --                                    & 1,134,350 \\
$\gamma$-progress & Curiosity   & --                                    & 226,870 \\
MAX               & Curiosity   & --                                    & 40,603,008 \\
NGU               & Curiosity   & --                                    & 449,063 \\
C-TeC             & Curiosity    & --                                    & 325,505 \\
RE3               & Entropy    & --                                    & 0 (fixed random encoder) \\
APT               & Entropy    & --                                    & 609,287 \\
Proto-RL          & Entropy    & --                                    & 507,136 \\
DIAYN             & Competency       & $\mathrm{Cat}(50)$ (per episode)       & 198,650 \\
VIC               & Competency       & $\mathrm{Cat}(12)$ (50 decisions)      & 227,852 \\
DADS              & Competency       & $\mathcal{U}([-1,1]^2)$ (100 decisions)  & 740,972 \\
rDADS             & Competency       & $\mathcal{U}([-1,1]^2)$ (100 decisions)  & 1,006,600 \\
VALOR             & Competency       & $\mathrm{Cat}(12)$ (50 decisions)      & 1,067,020 \\
CSF               & Competency       & $\mathcal{U}(S^3)$ (per episode)       & 1,372,164 \\
Forward empowerment        & Competency & --                           & 617,420 \\
Empowerment gain           & Competency & --                           & 3,124,150 \\
Potential empowerment      & Competency & --                           & 620,470 \\
Process empowerment        & Competency & --                           & 620,470 \\
\bottomrule
\end{tabular}
\caption{Intrinsic drive methods. Parameter counts exclude frozen targets, lagged copies, and memories.}
\label{tab:idm_summary}
\end{table}

\subsubsection*{Curiosity}
\label{appendix: curiosity}
\paragraph{ICM \citep{pathak2017icm}.} An encoder $g$ ($310\!\to\!256\!\to\!256$) is trained jointly with a forward model $f$ ($263\!\to\!256\!\to\!256\!\to\!256$) and an inverse model ($512\!\to\!256\!\to\!256\!\to\!7$), with loss $0.8\,\mathcal{L}_{\mathrm{inv}} + 0.2\,\mathcal{L}_{\mathrm{fwd}}$. The reward is $r_t = \tfrac{\eta}{2}\lVert f(g(\phi_t), a_t) - g(\phi_{t+1})\rVert^2$ with $\eta = 0.01$. Unlike the original, the encoder operates on the policy's features rather than learning from pixels end to end.

\paragraph{RND \citep{burda2019rnd}.} A predictor $\hat{f}$ regresses a frozen random target $f$ (both $310\!\to\!256\!\to\!256\!\to\!512$). The reward is the squared error $\lVert \hat{f}(\phi_t) - f(\phi_t)\rVert^2$, averaged over dimensions and divided by a running mean of the absolute error. The original normalizes by the standard deviation of intrinsic returns and uses separate intrinsic and extrinsic value heads with non-episodic intrinsic returns; here there is one value head and returns are episodic.

\paragraph{Disagreement \citep{pathak2019disag}.} Five forward models ($317\!\to\!256\!\to\!256\!\to\!310$), each trained with bootstrap-resampled mean squared error to predict $\phi_{t+1}$ from $(\phi_t, a_t)$. The reward is the variance of their predictions, averaged over dimensions, without normalization.

\paragraph{$\gamma$-progress \citep{kim2020gammaprogress}.} A forward model $f$ ($317\!\to\!256\!\to\!256\!\to\!310$) and a lagged copy $f_{\mathrm{old}}$, updated as an exponential moving average with rate $0.01$ per IDM update ($\gamma = 0.99$). The reward is the reduction in prediction error of $\phi_{t+1}$, $r_t = \mathcal{L}(f_{\mathrm{old}}) - \mathcal{L}(f)$, and can be negative.

\paragraph{MAX \citep{shyam2019max}.} Thirty-two diagonal-Gaussian forward models (four hidden layers of 512 units, swish activations, $\log\sigma^2$ bounded to $[-5, -1]$), trained by negative log-likelihood on bootstrap samples. The reward is the Jensen--R\'enyi (quadratic) divergence between the ensemble's predictions, with variance tempering $\lambda = 0.1$, clamped at zero. The original plans exploration inside the learned model; here the divergence is used as a model-free reward.

\paragraph{NGU \citep{badia2020ngu}.} The reward is $r_t = 0.3\, r^{\mathrm{epi}}_t \cdot \mathrm{clip}\!\left(1 + (e^{\mathrm{RND}}_t - \mu)/\sigma,\ 1,\ 5\right)$, where $e^{\mathrm{RND}}_t$ is an RND error with running mean $\mu$ and standard deviation $\sigma$. The episodic term is $r^{\mathrm{epi}}_t = 1/\big(\sqrt{\textstyle\sum_k K(e_t, e_k)} + 10^{-3}\big)$ over the 10 nearest neighbors in an episodic memory reset at each episode, set to zero above a similarity threshold of 8 as in the original, with kernel $K(x, y) = \epsilon/\big(\max(d^2(x,y)/\bar{d}^2 - \xi,\ 0) + \epsilon\big)$, $\epsilon = 10^{-4}$, $\xi = 8\times10^{-3}$. Embeddings $e_t$ ($310\!\to\!256\!\to\!256\!\to\!32$) are trained with an inverse model ($64\!\to\!256\!\to\!256\!\to\!7$); the RND networks are $310\!\to\!256\!\to\!256\!\to\!256$. We use a single mixing coefficient rather than the original family of policies.

\paragraph{C-TeC \citep{mohamed2026ctec}.} A state--action encoder $\psi$ ($317\!\to\!256\!\to\!256\!\to\!64$) and a future-state encoder $\nu$ ($310\!\to\!256\!\to\!256\!\to\!64$), both with $\ell_2$-normalized outputs, are trained with an InfoNCE loss using negative $\ell_1$ distance as the logit, 256 in-batch negatives, and a learned temperature. The reward is $r_t = \lVert \psi(\phi_t, a_t) - \nu(\phi_{t+1+\Delta})\rVert_1$ for a single offset $\Delta$ drawn from a geometric distribution ($\gamma = 0.99$) truncated to the episode. Rewards and advantages are recomputed at the end of each rollout.

\subsubsection*{Entropy}

\paragraph{RE3 \citep{seo2021re3}.} A fixed, randomly initialized encoder $h$ ($310\!\to\!256\!\to\!256\!\to\!50$) embeds $\phi_t$. The reward is the distance to the 53rd nearest neighbor in a memory of 2,048 embeddings sampled from recent training data (re-encoded every rollout), divided by its running RMS. The original applies a random convolutional encoder to pixels with a $\log(1 + d)$ reward and a decaying coefficient; here the random encoder acts on the policy's features, with no log and a constant coefficient of 1.

\paragraph{APT \citep{liu2021apt}.} An encoder $g$ ($310\!\to\!256\!\to\!256\!\to\!256$) is trained with forward ($263\!\to\!256\!\to\!256\!\to\!256$) and inverse ($512\!\to\!256\!\to\!256\!\to\!7$) losses. The reward is $r_t = \log\!\big(1 + \max(\bar{d}_{16}/\bar{d} - 5\times10^{-4},\ 0)\big)$, where $\bar{d}_{16}$ is the mean distance from the $\ell_2$-normalized embedding of $\phi_{t+1}$ to its 16 nearest neighbors in a memory of 2,048 embeddings (re-embedded after every update), and $\bar{d}$ is a running mean distance.

\paragraph{Proto-RL \citep{yarats2021protorl}.} An encoder ($310\!\to\!256\!\to\!256\!\to\!256$), projector ($256\!\to\!256\!\to\!128$), and predictor ($128\!\to\!512\!\to\!128$) are trained with a swapped-prediction loss against 512 prototypes (Sinkhorn with 3 iterations, temperature 0.1) and an exponential-moving-average target network (rate 0.95). The reward is the distance from the embedding of $\phi_{t+1}$ to its third-nearest neighbor in a queue of 2,048 candidate embeddings. Unlike the original, the representation is trained online alongside PPO rather than in a separate pre-training phase, without data augmentation, and the queue is not re-embedded when the encoder changes.

\subsubsection*{Competency}

\paragraph{DIAYN \citep{eysenbach2019diayn}.} Skills are one-hot, $z \sim \mathrm{Cat}(50)$ uniform, sampled once per training episode. A discriminator $q(z \mid \phi)$ ($310\!\to\!300\!\to\!300\!\to\!50$) is trained with cross-entropy (batch 128), and the reward is $r_t = \log q(z \mid \phi_t) - \log p(z) = \log q(z \mid \phi_t) + \log 50$. The original maximizes action entropy through SAC; here the only action-entropy term is PPO's entropy bonus (coefficient 0.001), and the discriminator scores the pre-transition features $\phi_t$.

\paragraph{VIC \citep{gregor2017vic}.} Skills $z \sim \mathrm{Cat}(12)$, uniform, fixed over 50-decision segments. A posterior $q(z \mid \phi_0, \phi_T)$ ($620\!\to\!256\!\to\!256\!\to\!12$) is trained with cross-entropy on 512 completed segments, and the reward $\log q(z \mid \phi_0, \phi_T) + \log 12$ is given only at a segment's final step. The prior is fixed and segments have fixed length rather than learned termination.

\paragraph{DADS \citep{sharma2020dads}.} Skills $z \sim \mathcal{U}([-1,1]^2)$, fixed over $100$-decisions or resampled at episode termination. A skill-dynamics model $q(\phi' \mid \phi, z)$, a single diagonal Gaussian ($312\!\to\!512\!\to\!512\!\to\!620$), predicts the next features. The reward is
\begin{equation}
r_t = \log q(\phi_{t+1} \mid \phi_t, z) - \log \frac{1}{L}\sum_{i=1}^{L} q(\phi_{t+1} \mid \phi_t, z_i), \qquad z_i \sim p(z),\ L = 100,
\end{equation}
divided by its running RMS. The original uses a mixture-of-experts dynamics model that predicts state differences; here the model is a single Gaussian over absolute next features.

\paragraph{rDADS.} A reverse formulation of DADS that infers the skill from transitions. The reward is $r_t = \log q_\psi(z \mid \phi_t, \phi_{t+1}) - \log q_\omega(z \mid \phi_t)$, with Gaussian posteriors $q_\psi$ ($620\!\to\!512\!\to\!512\!\to\!4$) and $q_\omega$ ($310\!\to\!512\!\to\!512\!\to\!4$) over $z \sim \mathcal{U}([-1,1]^2)$, trained by negative log-likelihood; the reward is divided by its running RMS.

\paragraph{VALOR \citep{achiam2018valor}.} Skills $z \sim \mathrm{Cat}(12)$, fixed over 50-decision segments. The decoder embeds the ten feature differences between eleven evenly spaced states of a segment ($310\!\to\!256\!\to\!256$ each) and maps their concatenation ($2560\!\to\!256\!\to\!512\!\to\!256\!\to\!12$) to a posterior over $z$. The reward $\log q(z \mid \tau) + \log 12$ is given at a segment's final step. 

\paragraph{CSF \citep{zheng2025csf}.} Skills are drawn uniformly from the unit sphere in $\mathbb{R}^4$, once per episode. A representation $f$ ($310\!\to\!1024\!\to\!1024\!\to\!4$) is trained to minimize $-\mathbb{E}\big[\Delta f^\top z - \xi \log \sum_{j=1}^{256} \exp(\Delta f^\top z_j)\big]$, where $\Delta f = f(\phi_{t+1}) - f(\phi_t)$, $\xi = 5$, and the negatives $z_j$ are drawn from the prior. The reward is $r_t = \Delta f^\top z$. The original trains a successor-feature critic; here PPO's value function is used.

\paragraph{Forward empowerment.} The reward is $r_t = \log q_\theta(\phi_{t+1} \mid \phi_t, a_t) - \log q_\omega(\phi_{t+1} \mid \phi_t)$, with diagonal-Gaussian models (two hidden layers of 256 units; the action is re-concatenated at every layer) trained by maximum likelihood. This estimates the one-step mutual information $I_\pi(A; \Phi' \mid \Phi)$ under the current policy, without maximizing over action distributions.

\paragraph{Potential empowerment.} Following \citet{klyubin2008emppot, mohamed2015empgainfwd, zhao2021empest}, the method learns a latent Gaussian channel. An encoder $f$ maps features to $y \in \mathbb{R}^{64}$, and $y' = f(\phi_{t+1})$ is modeled as Gaussian with a mean $\mu_a(y)$ formed from a drift term and a per-action gain, a shared learned log-variance $\ell$, and a decoder $64\!\to\!310$ for reconstruction (two hidden layers of 256 units). The loss is $\tfrac12\,\mathbb{E}\big[(y' - \mu_a)^2 e^{-\ell} + \ell\big] + \lVert \mathrm{dec}(f(\phi_t)) - \phi_t \rVert^2/310$, with gradients flowing through both $y'$ and $\mu_a$. The reward is the channel capacity $C(\phi_t) = \max_{p(a)} I(A; Y' \mid \phi_t)$ over the seven actions, computed with $32$ log-space Blahut--Arimoto iterations and $64$ antithetic samples.

\paragraph{Empowerment gain.} Uses the channel of potential empowerment with a larger network (five hidden layers of 400 units) and a slowly updated copy (exponential moving average with rate 0.01 per IDM update). The reward is the increase in estimated capacity, $r_t = C_{\mathrm{fast}}(\phi_t) - C_{\mathrm{slow}}(\phi_t)$, divided by its running RMS.

\paragraph{Process empowerment \citep{tiomkin2025empproc}.} Uses the channel of potential empowerment, with reward $r_t = \log q_{a_t}(y_{t+1}) - \log \sum_b \pi(b \mid h_t)\, q_b(y_{t+1})$, where $\pi(\cdot \mid h_t)$ are the policy's action probabilities from the pre-step LSTM state. The reward can be negative.

\subsection{Training Protocol}
\label{app:train_proto}

\paragraph{Environment.} We train each policy in a procedurally randomized FlyGym environment (Fig.~\ref{fig:fig_1_overview}a) that exposes the agent to variation in its visual, olfactory, and proprioceptive inputs. Each arena contains three odor sources and $12$ to $50$ obstacles, regenerated at every episode. Odor sources are placed uniformly in $[-35, 35]^2$\,mm at a height of $2$\,mm, at least $15$\,mm from the spawn point and $12$\,mm from each other. Obstacles are placed uniformly in $[-45, 45]^2$\,mm, at least $4$\,mm from the spawn point and $5$\,mm from odor sources and from each other. The fly spawns at the origin with zero yaw. The odor sources and obstacles vary the visual and olfactory observations, and physical contact with obstacles further diversifies proprioceptive input. Episodes last $5$\,s ($50$ decisions) and terminate early on a physics error or non-finite state, when the body falls below a height threshold for five consecutive decisions, or when the fly leaves $|x|, |y| \le 60$\,mm.

\paragraph{Optimization.} Policies are trained with Recurrent PPO \citep{raffin2021stablebaselines3, schulman2017ppo} in $2,048$ parallel environments, each with a distinct random seed, using the settings in Table~\ref{tab:ppo_params} and generalized advantage estimation \citep{schulman2016gae}. Each method is trained for the same budget of $50$ million policy decisions, summed across environments ($5\times10^{6}$\,s of simulated experience, approximately $244$ PPO updates). The scalar reward used by PPO is supplied entirely by the intrinsic drive module, so differences between trained policies arise from the intrinsic objective rather than from external task reward. LSTM states are reset at episode starts and carried across rollout boundaries; backpropagation through time runs over sequences split at episode boundaries (at most $50$ decisions). Truncated episodes are bootstrapped with the value of the final observation, and optimization epochs stop early when the approximate KL divergence exceeds $1.5$ times the target. Each method is trained once, with seed $0$. We evaluate the checkpoint at the end of the training budget, without selecting checkpoints on any evaluation metric.

\paragraph{Compute and software.} Each run used one GPU (NVIDIA L40S or comparable) and 64 CPU cores. Throughput ranged from roughly 80-129 policy decisions per second depending on the method. Recorded training wall time was 108-175 hours per method (Table~\ref{tab:train_wallclock}). These times exclude setup, the final PPO update, interrupted work without a rollout log, and time between jobs, so they are lower bounds on full job time. Software: Python 3.12, FlyGym 2.1.0, MuJoCo 3.9.0 with MuJoCo-Warp 3.9.0.1, Warp 1.14.0, Stable-Baselines3 and sb3-contrib 2.8.0, PyTorch, and JAX 0.4.33.

\begin{table}[t]
\centering
\small
\begin{tabular}{lrlrlr}
\toprule
Method & Hours & Decisions/s & Method & Hours & Decisions/s \\
\midrule
ICM               & 124.1 & 112.3 & DIAYN                 & 145.9 & 95.5 \\
RND               & 144.2 & 96.6  & VIC                   & 127.2 & 109.6 \\
Disagreement      & 121.1 & 115.1 & DADS                  & 155.1 & 89.9 \\
$\gamma$-progress & 162.8 & 85.6  & rDADS                 & 107.7 & 129.4 \\
MAX               & 139.4 & 100.0 & VALOR                 & 125.1 & 111.4 \\
NGU               & 175.3 & 79.5  & CSF                   & 146.1 & 95.4 \\
C-TeC             & 118.1 & 118.0 & Forward empowerment   & 166.2 & 83.9 \\
RE3               & 157.8 & 88.3  & Empowerment gain      & 156.8 & 88.9 \\
APT               & 150.0 & 92.9  & Potential empowerment & 118.0 & 118.1 \\
Proto-RL          & 150.3 & 92.7  & Process empowerment   & 108.8 & 128.1 \\
\bottomrule
\end{tabular}
\caption{Recorded training time for the IDM-trained policies. Hours sum the final rollout-end timer reading from each process, rounded to 0.1 hour, and are lower bounds on full job time. Decisions/s is $50{,}176{,}000$ divided by that summed time.}
\label{tab:train_wallclock}
\end{table}

\begin{table}[t]
\centering
\small
\begin{tabular}{ll}
\toprule
Setting & Value \\
\midrule
\multicolumn{2}{l}{\emph{Recurrent PPO}} \\
Parallel environments & 2,048 \\
Decisions per environment per rollout & 100 \\
Transitions per rollout & 204,800 \\
Minibatch size & 4,096 (50 minibatches per epoch) \\
Epochs per rollout & 10 \\
Learning rate & $3\times10^{-4}$ \\
Discount $\gamma$ & 0.99 \\
GAE $\lambda$ & 0.95 \\
Clip range & 0.2 \\
Target KL & 0.03 \\
Entropy coefficient & 0.001 \\
Value-function coefficient & 0.5 \\
Maximum gradient norm & 0.5 \\
Training budget & $50\times10^{6}$ policy decisions ($\approx 244$ rollouts) \\
Observation normalization & Odor and proprioception, clipped at 10 \\
\midrule
\multicolumn{2}{l}{\emph{Simulation}} \\
Physics timestep & $10^{-4}$\,s \\
Decision interval & $0.1$\,s (1,000 physics steps) \\
Episode length & 5\,s (50 decisions) \\
Extrinsic reward scale & 0 \\
Intrinsic reward scale & 1 \\
\bottomrule
\end{tabular}
\caption{PPO training and environment hyperparameters.}
\label{tab:ppo_params}
\end{table}

\subsection{Tethered-Ball Evaluation}
\label{app:eval}

\paragraph{Assay.} In the biological experiment, head-fixed flies walk on a $9$\,mm air-suspended foam ball in darkness, and FicTrac \citep{moore2014fictrac} recovers the ball's angular and forward velocity from an infrared-visible painted pattern \citep{brezovec2023volitional}. In simulation, we read these velocities directly from the ball. The ball is a $9$\,mm-diameter sphere with the density of the experimental foam ($0.240$\,mg/mm$^3$; mass $\approx 0.092$\,g) \citep{generalplastics_fr4615} on an undamped ball joint. The fly's body root is fixed in place above the ball, and only the tibiae and tarsi contact it (friction coefficient 2.0). Because the root is fixed, the body linear- and angular-velocity entries of the proprioceptive observation are zero. Forward velocity is $v = -R\,\omega^{\mathrm{ball}}_y$ (mm/s), where $R$ is the ball radius, and angular velocity is the ball's yaw rate (deg/s, positive for left turns); both are recorded at 50\,Hz after an unrecorded settling period.

\paragraph{Sensory masking.} To match the darkness and control for odor in the biological experiment, the raw visual and olfactory observations are set to zero.

\paragraph{Rollouts.} Each frozen policy runs nine 30-minute rollouts (18,000 decisions each), matching the nine flies and 270 minutes of the biological dataset. Rollouts differ in their initial CPG phases (seeds 0--8) and in the actions sampled from the policy; actions are sampled rather than chosen greedily. The LSTM state is initialized once and never reset. For DIAYN and CSF, one skill is sampled per rollout and held for the full 30 minutes; VIC and VALOR resample every 5\,s (50 decisions), and DADS and rDADS every 10\,s (100 decisions).

\paragraph{Recorded activations.} At every decision we record the input embedding $\phi_t$, the actor and critic LSTM hidden states, and the policy and value MLP hidden representations. Activations are averaged over consecutive $0.532$\,s bins from rollout onset, matching the fly volume rate (1.88\,Hz), giving 3,384 bins per rollout.

\subsection{Controls}
\label{app:controls}

\paragraph{Random initialization.} Each control policy is a random initialization of a source policy so the LSTM input has the same size as in the trained policy. Random-DIAYN, Random-CSF, Random-VIC (the reference for VIC and VALOR), and Random-DADS (the reference for DADS and rDADS) retain the corresponding skill input; Random-nonskill is derived from a run of a method without a skill input. Controls are evaluated exactly as trained policies.

\paragraph{Extrinsic reward.} The extrinsic baseline uses the same architecture and PPO settings, but is trained on an odor-seeking task instead of an intrinsic reward. Each arena has one odor source at $x \in [20, 30]$\,mm, $y \in [-6, 6]$\,mm and 3 to 5 obstacles, and the reward is $r_t = (d_{t-1} - d_t) - 0.01 + 10\,\mathbf{1}_{\{d_t < 2\,\mathrm{mm}\}}$, where $d_t$ is the distance to the source. Episodes terminate on reaching the source, and reward normalization is enabled. The baseline was trained for $10$M decision steps, by which it had saturated the task success rate. This final checkpoint was used for evaluation.

\subsection{Biological Data}
\label{app:bio}

We use the whole-brain recordings of \citet{brezovec2023volitional}, released on the DANDI Archive\citep{brezovec2024mapping, brezovec2024dandi}: nine flies (IDs 087, 089, 094, 097, 098, 099, 100, 101, 105) and the processed functional GCaMP6f signal. Brain regions are defined by the JFRC2010 \citep{jenett2012gal4} anatomical template, warped to the JRC2018F \citep{bogovic2020unbiased} template with ANTs SyN \citep{avants2011ants, avants2009ants}. The supercluster that \citet{brezovec2023volitional} identify as the pre-motor center is not available in the public release, so we use the whole anatomically defined IPS in each hemisphere.

Turns are re-detected from the FicTrac yaw rate as described in Appendix~\ref{app:behavior_analysis}, discarding the first and last 40\,s of each 30-minute session. This yields 3,786 turns (1,362 left, 2,424 right), matching the count reported by \citet{brezovec2023volitional}. Turn direction is strongly imbalanced within some flies (Table~\ref{tab:fly_turns}); \citet{brezovec2023volitional} attribute part of the per-fly rotational skew to small asymmetries in how each fly was mounted on the ball. Turns are pooled across flies. The stationary subset consists of the 27 left and 27 right turns with the least movement (mean $|v|$ from $-15$ to $-1$\,s), pooled across flies.

\begin{table}[t]
\centering
\small
\begin{tabular}{lrrr}
\toprule
Fly & $N_L$ & $N_R$ & $\log(N_R/N_L)$ \\
\midrule
087 & 222 & 519 & $0.85$ \\
089 & 38 & 33 & $-0.14$ \\
094 & 140 & 2 & $-4.25$ \\
097 & 47 & 105 & $0.80$ \\
098 & 178 & 291 & $0.49$ \\
099 & 27 & 543 & $3.00$ \\
100 & 364 & 601 & $0.50$ \\
101 & 276 & 219 & $-0.23$ \\
105 & 70 & 111 & $0.46$ \\
\midrule
Total & 1,362 & 2,424 & $0.17 \pm 1.90$ (mean $\pm$ s.d.) \\
\bottomrule
\end{tabular}
\caption{Turns per fly in the biological dataset.}
\label{tab:fly_turns}
\end{table}

\subsection{Turn Detection and Behavioral Statistics}
\label{app:behavior_analysis}

For agents and flies, angular and forward velocity, sampled at 50\,Hz, are smoothed with a Savitzky--Golay filter \citep{savitzky1964smoothing} (25 samples, i.e.\ $0.5$\,s; polynomial order 3). Turns are local maxima of $|\omega|$ exceeding $200^\circ$/s, following \citet{brezovec2023volitional}; the turn time is the time of the peak. For flies, peaks in the first and last 40\,s of each session are discarded. To compare behavior, we compute the number of left and right turns, $N_L$ and $N_R$, for each rollout or fly, and the turning bias $\log(N_R/N_L)$. Turning bias is reported only when $N_L > 0$ and $N_R > 0$; otherwise it is undefined, and such rollouts are omitted from bias summaries but retained in turn counts.

\subsection{Decoding}
\label{app:analysis}

\paragraph{Features.} For each turn, features are averaged over a window relative to the turn: \beforeturn{} (before the turn) or \duringturn{} (during the turn). For agents, the average is over the $0.532$\,s activation bins whose left edge falls within the window. We record five internal network representations: the learned observation embedding, the actor LSTM hidden state, the critic LSTM hidden state, the policy MLP hidden representation and the value MLP hidden representation.  The network decoding (neural prospective coding) analysis evaluates seven candidate feature sets: the actor LSTM, critic LSTM, policy MLP and value MLP individually, and three concatenations of layers:  actor LSTM with policy MLP (\textsc{Actor}), critic LSTM with value MLP (\textsc{Critic}), and all five recorded internal representations (\textsc{All}). The learned observation embedding is included in \textsc{All}. As a separate input-level control, we decode from the raw proprioceptive observations with vision and odor set to zero. For flies, features are the window-averaged activity of every voxel in the left and right IPS, concatenated; unlike \citet{brezovec2023volitional}, we neither restrict the analysis to the pre-motor supercluster nor use the interhemispheric difference. Behavioral features are the window means of angular and forward velocity.

\paragraph{Classifier.} We use logistic regression with an elastic-net penalty \citep{zou2005elasticnet, pedregosa2011scikit} and the saga solver \citep{defazio2014sagafastincrementalgradient}. Hyperparameters were chosen by nested cross-validation on the fly IPS data over $C \in \{10^{-4}, 10^{-3}, \dots, 10^{4}\}$ and $\ell_1$ ratio $\in \{0.25, 0.5, 0.75\}$, and the most frequently selected configuration ($C = 0.1$, $\ell_1$ ratio $0.75$) is used for all decoders of agents and flies. Features are standardized with statistics from the training fold only.
\paragraph{Evaluation.} Accuracy is estimated by stratified $k$-fold cross-validation over individual turns, with $k = \min(5, N_{\min})$, where $N_{\min}$ is the smaller class count; methods with fewer than two turns in either class are not scored. We refer to the balanced accuracy \citep{brodersen2010balanced} of a neural or behavioral decoder as its prospective coding,
\begin{equation}
\mathrm{BAcc} = \frac{1}{2}\left(\frac{TP_L}{N_L} + \frac{TP_R}{N_R}\right),
\end{equation}
where $TP_L$ and $TP_R$ are the numbers of correctly classified left and right turns and $N_L$ and $N_R$ are the numbers of left and right turns. We report the mean $\pm$ s.d. across folds. For each method, the selected layer is the candidate layer with the highest before-turn prospective coding, and the same layer is used for the during-turn window.

\subsection{Representational Analysis}
\label{app:rep_analysis}
\label{app:pca}

Because the policy has no anatomical analogue of the fly's bilaterally matched superclusters, we cannot reproduce the spatial pairing analysis of \citet{brezovec2023volitional} directly. Instead, we use PCA to identify low-dimensional axes of model activity and ask whether pairs of components show the same functional relationship to behavior: opposing activity associated with angular velocity and shared activity associated with forward velocity. This is a functional comparison of representational geometry, not an anatomical correspondence between model components and fly hemispheres.

For DIAYN and Random-DIAYN, we fit a 10-component PCA (randomized SVD, seed 0) to the actor LSTM activations pooled across all nine rollouts ($9 \times 3{,}384$ samples; units centered but not standardized), and scale each component to unit variance. Angular and forward velocity are interpolated to the activation timestamps. For every unordered pair $(i, j)$ of components, we compute
\begin{equation}
S_{ij} = \left| r\!\left(\mathrm{PC}_i - \mathrm{PC}_j,\ \omega\right) \right| + \left| r\!\left(\mathrm{PC}_i + \mathrm{PC}_j,\ v\right) \right|,
\end{equation}
where $r$ is the Pearson correlation over all timepoints, and select the maximizing pair, breaking ties by the smaller of the two correlations. The pair is oriented so that the difference increases with angular velocity. The same search is applied to the randomly initialized policy as a reference. The selected pairs are PCs 2 and 8 for DIAYN and PCs 5 and 10 for Random-DIAYN (Fig.~\ref{fig:rep_analysis}). The biological panel (Fig.~\ref{fig:rep_analysis}a) is reconstructed from Fig.~1e--f of \citet{brezovec2023volitional}.

\section{Extended Results}
\label{app:results}

Table~\ref{tab:a11-extended} reports turn statistics and prospective coding for all 20 intrinsic drive methods, the randomly initialized controls, the extrinsic-reward control, and the flies. Table~\ref{tab:a11-layers} gives the before-turn prospective coding of every candidate layer; ``Obs'' denotes the proprioceptive observation.

\begin{table}[t]
\centering\scriptsize\setlength{\tabcolsep}{3pt}
\resizebox{\textwidth}{!}{%
\begin{tabular}{lrrcclccrccl}
\toprule
Method & $N_L$ & $N_R$ & Turns/min & Turn bias $\log(N_R/N_L)$ & Sel.\ layer & Layer acc.\ ($-15..-8$\,s) & Behav.\ acc.\ ($-15..-8$\,s) & Margin & Internal ($\pm0.4$\,s) & Behav.\ ($\pm0.4$\,s) & Unscored reason \\
\midrule
ICM & 0 & 14997 & 55.55 $\pm$ 1.30 & -- & -- & -- & -- & -- & -- & -- & only right turns \\
RND & 0 & 0 & 0.00 $\pm$ 0.00 & -- & -- & -- & -- & -- & -- & -- & no turns \\
Disagreement & 0 & 62288 & 230.70 $\pm$ 12.13 & -- & -- & -- & -- & -- & -- & -- & only right turns \\
$\gamma$-progress & 10320 & 295 & 39.32 $\pm$ 9.03 & -3.78 $\pm$ 0.96 & A-LSTM+Pol & 0.603 $\pm$ 0.022 & 0.724 $\pm$ 0.012 & -0.122 & 0.864 $\pm$ 0.028 & 1.000 $\pm$ 0.000 & -- \\
RE3 & 179 & 5379 & 20.59 $\pm$ 1.90 & 3.47 $\pm$ 0.49 & A-LSTM (8 tied)$^{\ddagger}$ & 0.500 $\pm$ 0.000 & 0.531 $\pm$ 0.050 & -0.031 & 0.756 $\pm$ 0.019 & 1.000 $\pm$ 0.000 & -- \\
APT & 252 & 534 & 2.91 $\pm$ 0.51 & 0.79 $\pm$ 0.27 & All & 0.516 $\pm$ 0.018 & 0.534 $\pm$ 0.031 & -0.019 & 0.964 $\pm$ 0.013 & 1.000 $\pm$ 0.000 & -- \\
Proto-RL & 0 & 0 & 0.00 $\pm$ 0.00 & -- & -- & -- & -- & -- & -- & -- & no turns \\
MAX & 4 & 18486 & 68.48 $\pm$ 33.12 & 5.14 $\pm$ 0.00$^{\dagger}$(1/9) & A-LSTM (8 tied)$^{\ddagger}$ & 0.500 $\pm$ 0.000 & 0.922 $\pm$ 0.059 & -0.422 & 0.500 $\pm$ 0.000 & 0.985 $\pm$ 0.025 & -- \\
Emp. Forward & 10212 & 4 & 37.84 $\pm$ 7.35 & -6.88 $\pm$ 0.14$^{\dagger}$(4/9) & A-LSTM (8 tied)$^{\ddagger}$ & 0.500 $\pm$ 0.000 & 0.341 $\pm$ 0.167 & +0.159 & 0.500 $\pm$ 0.000 & 0.983 $\pm$ 0.027 & -- \\
Emp. Gain & 0 & 134808 & 499.29 $\pm$ 6.28 & -- & -- & -- & -- & -- & -- & -- & only right turns \\
Emp. Potential & 63 & 4850 & 18.20 $\pm$ 0.85 & 4.39 $\pm$ 0.31 & A-LSTM (8 tied)$^{\ddagger}$ & 0.500 $\pm$ 0.000 & 0.508 $\pm$ 0.021 & -0.008 & 0.548 $\pm$ 0.030 & 1.000 $\pm$ 0.000 & -- \\
Emp. Process & 0 & 0 & 0.00 $\pm$ 0.00 & -- & -- & -- & -- & -- & -- & -- & no turns \\
DIAYN & 894 & 785 & 6.22 $\pm$ 2.04 & -0.37 $\pm$ 2.27 & A-LSTM & 0.821 $\pm$ 0.009 & 0.747 $\pm$ 0.014 & +0.074 & 0.985 $\pm$ 0.005 & 1.000 $\pm$ 0.000 & -- \\
VIC & 0 & 160 & 0.59 $\pm$ 0.41 & -- & -- & -- & -- & -- & -- & -- & only right turns \\
DADS & 23 & 0 & 0.09 $\pm$ 0.05 & -- & -- & -- & -- & -- & -- & -- & only left turns \\
rDADS & 79348 & 0 & 293.88 $\pm$ 13.52 & -- & -- & -- & -- & -- & -- & -- & only left turns \\
VALOR & 125 & 19 & 0.53 $\pm$ 0.25 & -1.78 $\pm$ 1.11 & A-LSTM (8 tied)$^{\ddagger}$ & 0.500 $\pm$ 0.000 & 0.496 $\pm$ 0.095 & +0.004 & 0.875 $\pm$ 0.112 & 1.000 $\pm$ 0.000 & -- \\
CSF & 1830 & 280 & 7.81 $\pm$ 3.79 & -1.84 $\pm$ 1.03 & All & 0.527 $\pm$ 0.008 & 0.590 $\pm$ 0.025 & -0.063 & 0.921 $\pm$ 0.014 & 1.000 $\pm$ 0.000 & -- \\
NGU & 76 & 121 & 0.73 $\pm$ 0.70 & 0.38 $\pm$ 1.46$^{\dagger}$(6/9) & A-LSTM+Pol & 0.654 $\pm$ 0.018 & 0.663 $\pm$ 0.094 & -0.009 & 0.896 $\pm$ 0.110 & 1.000 $\pm$ 0.000 & -- \\
C-TeC & 6703 & 37 & 24.96 $\pm$ 1.29 & -5.52 $\pm$ 0.93 & A-LSTM (8 tied)$^{\ddagger}$ & 0.500 $\pm$ 0.000 & 0.578 $\pm$ 0.077 & -0.078 & 0.767 $\pm$ 0.045 & 1.000 $\pm$ 0.000 & -- \\
\midrule
Random-DIAYN & 195 & 189 & 1.42 $\pm$ 0.23 & -0.05 $\pm$ 0.34 & Value & 0.539 $\pm$ 0.049 & 0.523 $\pm$ 0.034 & +0.016 & 0.812 $\pm$ 0.035 & 1.000 $\pm$ 0.000 & -- \\
Random-CSF & 211 & 181 & 1.45 $\pm$ 0.23 & -0.12 $\pm$ 0.38 & C-LSTM & 0.498 $\pm$ 0.039 & 0.552 $\pm$ 0.036 & -0.054 & 0.944 $\pm$ 0.011 & 1.000 $\pm$ 0.000 & -- \\
Random-DADS & 184 & 185 & 1.37 $\pm$ 0.21 & 0.02 $\pm$ 0.23 & C-LSTM & 0.521 $\pm$ 0.054 & 0.509 $\pm$ 0.020 & +0.011 & 0.957 $\pm$ 0.010 & 1.000 $\pm$ 0.000 & -- \\
Random-VIC & 178 & 183 & 1.34 $\pm$ 0.24 & -0.00 $\pm$ 0.33 & C-LSTM & 0.558 $\pm$ 0.036 & 0.471 $\pm$ 0.031 & +0.087 & 0.909 $\pm$ 0.040 & 1.000 $\pm$ 0.000 & -- \\
Random-nonskill & 194 & 183 & 1.40 $\pm$ 0.15 & -0.06 $\pm$ 0.39 & Value & 0.512 $\pm$ 0.076 & 0.483 $\pm$ 0.031 & +0.029 & 0.870 $\pm$ 0.041 & 1.000 $\pm$ 0.000 & -- \\
Extrinsic & 0 & 10541 & 39.04 $\pm$ 19.43 & -- & -- & -- & -- & -- & -- & -- & only right turns \\
\midrule
Fly (IPS) & 1362 & 2424 & -- & 0.17 $\pm$ 1.90 & IPS & 0.665 $\pm$ 0.014 & 0.605 $\pm$ 0.009 & +0.060 & 0.721 $\pm$ 0.018 & 0.955 $\pm$ 0.016 & -- \\
\bottomrule
\end{tabular}}
\caption{Extended per-method results. Accuracy is balanced accuracy; uncertainty is fold s.d. where shown. $\dagger$ Turning bias averaged over only the indicated number of rollouts; the remaining rollouts turned in only one direction (or not at all), so their bias is undefined. $N_L$ and $N_R$ are summed over all nine rollouts. $\ddagger$ marks a tie at maximum accuracy.}
\label{tab:a11-extended}
\end{table}

\begin{table}[t]
\centering\scriptsize\setlength{\tabcolsep}{3pt}
\begin{tabular}{lcccccccc}
\toprule
Method & A-LSTM & C-LSTM & Pol & Val & Actor & Critic & All & Raw obs. \\
\midrule
ICM & -- & -- & -- & -- & -- & -- & -- & -- \\
RND & -- & -- & -- & -- & -- & -- & -- & -- \\
Disagreement & -- & -- & -- & -- & -- & -- & -- & -- \\
$\gamma$-progress & 0.579 & 0.529 & 0.592 & 0.500 & \textbf{0.603} & 0.541 & 0.584 & 0.508 \\
RE3 & \textbf{0.500} & 0.500 & 0.500 & 0.500 & 0.500 & 0.500 & 0.500 & 0.500 \\
APT & 0.500 & 0.503 & 0.500 & 0.500 & 0.498 & 0.502 & \textbf{0.516} & 0.505 \\
Proto-RL & -- & -- & -- & -- & -- & -- & -- & -- \\
MAX & \textbf{0.500} & 0.500 & 0.500 & 0.500 & 0.500 & 0.500 & 0.500 & 0.500 \\
Emp. Forward & \textbf{0.500} & 0.500 & 0.500 & 0.500 & 0.500 & 0.500 & 0.500 & 0.500 \\
Emp. Gain & -- & -- & -- & -- & -- & -- & -- & -- \\
Emp. Potential & \textbf{0.500} & 0.500 & 0.500 & 0.500 & 0.500 & 0.500 & 0.500 & 0.500 \\
Emp. Process & -- & -- & -- & -- & -- & -- & -- & -- \\
DIAYN & \textbf{0.821} & 0.814 & 0.819 & 0.813 & 0.820 & 0.816 & 0.820 & 0.791 \\
VIC & -- & -- & -- & -- & -- & -- & -- & -- \\
DADS & -- & -- & -- & -- & -- & -- & -- & -- \\
rDADS & -- & -- & -- & -- & -- & -- & -- & -- \\
VALOR & \textbf{0.500} & 0.500 & 0.500 & 0.500 & 0.500 & 0.500 & 0.500 & 0.500 \\
CSF & 0.512 & 0.507 & 0.503 & 0.504 & 0.514 & 0.507 & \textbf{0.527} & 0.500 \\
NGU & 0.645 & 0.605 & 0.621 & 0.653 & \textbf{0.654}$^{\dagger}$ & 0.598 & 0.633 & 0.672 \\
C-TeC & \textbf{0.500} & 0.500 & 0.500 & 0.500 & 0.500 & 0.500 & 0.500 & 0.500 \\
\midrule
Random-DIAYN & 0.516 & 0.506 & 0.521 & \textbf{0.539}$^{\dagger}$ & 0.513 & 0.534 & 0.529 & 0.545 \\
Random-CSF & 0.474 & \textbf{0.498} & 0.496 & 0.477 & 0.477 & 0.485 & 0.489 & 0.478 \\
Random-DADS & 0.502 & \textbf{0.521} & 0.502 & 0.496 & 0.488 & 0.505 & 0.483 & 0.498 \\
Random-VIC & 0.519 & \textbf{0.558}$^{\dagger}$ & 0.520 & 0.541 & 0.508 & 0.545 & 0.503 & 0.581 \\
Random-nonskill & 0.468 & 0.472 & 0.503 & \textbf{0.512}$^{\dagger}$ & 0.485 & 0.480 & 0.510 & 0.520 \\
Extrinsic & -- & -- & -- & -- & -- & -- & -- & -- \\
\bottomrule
\end{tabular}
\caption{
Early-window (\beforeturn{}) balanced decoding accuracy for each candidate network feature set and the raw-observation control. A-LSTM and C-LSTM denote the actor and critic LSTM hidden states; Pol and Val denote the policy- and value-MLP hidden representations; Actor concatenates A-LSTM+Pol; Critic concatenates C-LSTM+Val; and All concatenates all five recorded internal representations, including the learned observation embedding. Raw obs. denotes the proprioceptive input with vision and odor set to zero and is distinct from the learned observation embedding. Bold denotes the selected network representation (argmax over the seven network feature sets only).$\dagger$ indicates that the raw-observation control achieved higher decoding accuracy than the selected internal representation. -- indicates that decoding was not scored because fewer than two bouts were available in one class. }
\label{tab:a11-layers}
\end{table}

\end{document}